\newcommand{\be}{\begin{equation}}
\newcommand{\ee}{\end{equation}}
\newcommand{\bea}{\begin{eqnarray}}
\newcommand{\eea}{\end{eqnarray}}
\newcommand{\ba}{\begin{array}}
\newcommand{\ea}{\end{array}}

\newcommand{\al}{\alpha}
\newcommand{\ga}{\gamma}
\newcommand{\Ga}{\Gamma}

\newcommand{\ka}{\kappa}
\newcommand{\de}{\delta}
\newcommand{\vphi}{\varphi}

\newcommand{\si}{\sigma}
\newcommand{\la}{\lambda}

\newcommand{\Om}{\Omega}
\newcommand{\ze}{\zeta}
\newcommand{\De}{\Delta}

\newcommand{\tha}{\theta}

\newcommand{\tphi}{\tilde{\phi}}
\newcommand{\tX}{\tilde{X}}

\newcommand{\tr}{{\rm tr}}
\newcommand{\Tr}{{\rm Tr}}

\newcommand{\D}{{\rm d}}
\newcommand{\pa}{\partial}
\newcommand{\rar}{\rightarrow}

\newcommand{\non}{\nonumber}
\newcommand{\we}{\wedge}
\newcommand{\cL}{\mathcal{L}}
\newcommand{\cD}{\mathcal{D}}
\newcommand{\cO}{\mathcal{O}}
\newcommand{\half}{\mbox{$\frac{1}{2}$}}
\newcommand{\B}{|B\rangle}
\newcommand{\BF}{|B(F)\rangle}
\newcommand{\ts}{\textstyle}

\documentclass[12pt]{article}

\usepackage{amsfonts}

\begin{document}
 
\begin{flushright}
 BRX-TH-494 \\ {\tt hep-th/0107185}
\end{flushright}
\vspace{1mm}
\begin{center}{\bf\Large\sf Derivative corrections to the D-brane Born-Infeld action: \\  non-geodesic embeddings and the Seiberg-Witten map}
\end{center}
\vskip 3mm
\begin{center}
Niclas Wyllard\footnote{Research supported by DOE grant DE-FG02-92ER40706.} \vspace{5mm}\\
{\em  Martin Fischer School of Physics, \\ Brandeis University, Waltham MA02454, USA\\ {\tt wyllard@brandeis.edu}}
\end{center}
 
\vskip 5mm
 
\begin{abstract}
 We dimensionally reduce the four-derivative corrections to the parity-conserving part of the D9-brane effective action involving all orders of the gauge field, to obtain corrections to the actions for the lower-dimensional D$p$-branes. These corrections involve the second fundamental form and correspond to a non-geodesic embedding of the D$p$-brane into (flat) ten-dimensional space. In addition, we  study the transformation of the corrections under the Seiberg-Witten map relating the ordinary and non-commutative theories. A speculative discussion about higher-order terms in the derivative expansion is also included.
\end{abstract}

\setcounter{equation}{0}
\section{Introduction}
D-branes \cite{Dai:1989} play a central role in current research in string theory. Many aspects of D-brane physics can be analyzed within the context of the effective field theory living on the D-brane. The action for this field theory is well understood for a single D-brane and slowly varying fields \cite{Fradkin:1985}. Some of the higher-order corrections in a derivative expansion are also known for a brane propagating in a flat supergravity background \cite{Andreev:1988,Wyllard:2000} and also for propagation in curved backgrounds \cite{Green:1997,Bachas:1999}.    
In this paper we study some aspects of the derivative corrections to the (abelian) effective action for a D-brane propagating in a flat background. 

In the next section we dimensionally reduce the recently determined expression for the four-derivative corrections to the parity-conserving Born-Infeld part of the D9-brane action, obtaining corrections to the actions for the lower-dimensional D$p$-branes.  We show that the part of these corrections involving only the transverse scalars can be written in terms of the second fundamental form as required on general grounds. The resulting expression is shown to agree with a previous result in the literature \cite{Bachas:1999}. 

In  section \ref{SWsec} the transformation properties of the derivative corrections to the D-brane action under the Seiberg-Witten map \cite{Seiberg:1999} are studied. Earlier work on this topic can be found in \cite{Okawa:1999,Terashima:2000,Okawa:2000,Cornalba:1999}. We begin by briefly discussing the zeroth-order terms in the derivative expansion (including the fermions and the transverse scalars) and their invariance under the Seiberg-Witten map. We then introduce some general terminology which will facilitate the later analysis and discuss the properties of manifest invariants under the Seiberg-Witten map. These results are then applied to the recently proposed  expression  \cite{Andreev:2001} for the two-derivative corrections involving all orders in $F$ to the action for a D-brane in the bosonic string theory.  
We conclude the section by showing that the subset of the four-derivative corrections to the parity-conserving Born-Infeld part of the D9-brane effective action which gives rise to the corrections discussed in section \ref{sLow} upon dimensional reduction are invariant under the Seiberg-Witten map, and briefly discuss how the rest of the terms might modify this result. Recently, the Seiberg-Witten map has been constructed to all orders in the non-commutativity parameter \cite{Liu:2000}. 
We will not make use of these results in this paper, although we will comment briefly on them in later sections. Even more recently some aspects of the transformation properties of the derivative corrections to the D9-brane action under the Seiberg-Witten map have been studied \cite{Das:2001}. There is some overlap with our work; however, we work to all orders in the gauge field strength but to  first order in the non-commutativity parameter, whereas in \cite{Das:2001} the results are obtained by working to  all orders in the non-commutativity parameter but order by order in the gauge field. The two methods are thus complementary, although for some applications the approach in \cite{Das:2001} appears to be more powerful.

Finally, section \ref{sSpec} contains a  speculative discussion about derivative corrections beyond the fourth order. In particular, we propose a formula connecting the derivative corrections to the Wess-Zumino term to derivative corrections to the Born-Infeld part. A few tests of this relation are successfully performed.

\setcounter{equation}{0}
\section{D$p$-brane actions: non-geodesic embeddings} \label{sLow}

In \cite{Wyllard:2000} we calculated some corrections to the parity-violating Wess-Zumino term for the D9-brane action. These corrections were dimensionally reduced leading to corrections for the lower-dimensional D$p$-branes, and it was shown that the part of the D$p$-brane action involving only the transverse scalars obtained that way correctly reproduces previously known results in the literature \cite{Bachas:1999}. 
In this section we will discuss the form of the four-derivative corrections in lower dimensions obtained by dimensional reduction of the parity-conserving part. For simplicity we will set the gauge field in the lower dimension to zero, thus retaining only the transverse scalars. On general grounds one expects that the resulting action should be expressible in terms of the second fundamental form, and we will find that this is indeed the case. 

We will label 
ten-dimensional indices by $M,N,\ldots$, world-volume indices by
$\mu,\nu,\ldots$, and indices for the transverse (normal) directions $i,j,\ldots$. There is a general theory describing the
embedding of a D$p$-brane into flat ten-dimensional space. The embedding is described by the objects $\pa_{\mu}Y^{M}$, which span a
local frame of the tangent bundle and the  $\xi^i_M$ which play a similar
role for the normal bundle (see e.g.\ \cite{Bachas:1999} for further details). World-volume indices are raised and lowered
with the induced metric $g_{\mu\nu} =
\de_{MN}\pa_{\mu}Y^{N}\pa_{\nu}Y^{M}$. Transverse indices are raised and
lowered with $\de_{ij}$ and ten-dimensional indices with $\de_{MN}$. The fundamental quantity describing the embedding is the
second fundamental form, $\Om^{M}_{\mu\nu}$, defined as
the covariant derivative of $\pa_{\mu}Y^{M}$, i.e.\ 
\be \label{Om}
\Om^{M}_{\mu\nu} = D_{\mu}\pa_{\nu}Y^{M} = \pa_{\mu}\pa_{\nu}Y^{M} -
{\Ga_{\!\!\mathrm{T}}}^{\la}_{\mu\nu}\pa_{\la}Y^{M} \,,
\ee
where ${\Ga_{\!\!\mathrm{T}}}^{\la}_{\mu\nu}$ is the connection constructed
from the induced metric. It is straightforward to show that $\Om^{M}_{\mu\nu} =
(\de^{M}_{L} - \pa_{\la} Y^{M}
g^{\la\rho}\pa_{\rho}Y^{K}\de_{KL})\pa_{\mu}\pa_{\nu}Y^{L} =:
P^{M}_{L}\pa_{\mu}\pa_{\nu}Y^{L}$, where $P^{M}_{L}$ is a projection
matrix, $P^{M}_{N} P^{N}_{L} = P^{M}_{L}$. The projection of $\Om^{M}{}_{\mu\nu}$ onto the tangent bundle vanishes, so it is sufficient to consider its projection onto the normal bundle,  
$\Om^i{}_{\mu\nu} := \xi^i_{M}\Om^{M}{}_{\mu\nu}$. The following relation is also useful, $\xi^i_{M}\xi^j_{N}\de_{ij} = P_{MN}$. When comparing with the dimensional reduction of the D9-brane action, we will use the static gauge in the lower dimension: $Y^{\mu} = x^{\mu}$ and $Y^i = X^i(x)$.

The parity-conserving part of the action for a D9-brane propagating in flat ten-dimensional space, correct up to four derivatives (order $\al'^2$ in our conventions\footnote{Our conventions are such that the (dimensionless) field strength $2\pi\al'F$ (which we will denote $F$) is of zeroth order. This makes the expansion in derivatives an expansion in $\al'$.}) is given by \cite{Wyllard:2000}
\bea \label{BIact}
S_{\mathrm{BI}} &=& - T_{9} \int \! \D^{10}x \sqrt{h}\Big( 1 + 
{\ts \frac{(2\pi\al')^2}{96}}\big[ -h^{\mu_4\mu_1}h^{\mu_2\mu_3}h^{\rho_4\rho_1}h^{\rho_2\rho_3}S_{\rho_1\rho_2\mu_1\mu_2}S_{\rho_3\rho_4\mu_3\mu_4}
 \non \\ && \qquad \qquad \qquad\qquad\qquad\; + \half
h^{\rho_4\rho_1}h^{\rho_2\rho_3}S_{\rho_1\rho_2}S_{\rho_3\rho_4}\big]\Big) \,,
\eea
where $h_{\mu\nu} = \de_{\mu\nu} + F_{\mu\nu}$,  $h^{\mu\la}$ is its inverse ($h^{\mu\nu}h_{\nu\la} = \de_\la^\mu$) and $h$ is short-hand for $\det(h_{\mu\nu})$. Furthermore,
\be \label{S}
S_{\rho_1\rho_2 \mu_1\mu_2} :=
\pa_{\rho_1}\pa_{\rho_2}F_{\mu_1\mu_2} + 2h^{\nu_1 \nu_2}\pa_{\rho_1}
F_{[\mu_1|\nu_1} \pa_{\rho_2|}F_{\mu_2]\nu_2}\,,
\ee
and 
 $S_{\rho_1\rho_2}:=h^{\mu_1\mu_2}S_{\rho_1\rho_2\mu_1\mu_2}$. Below we will also use the notation $h_S^{\mu\nu} = \frac{1}{2}[h^{\mu\nu} + h^{\nu\mu}]$ and $h_A^{\mu\nu} = \frac{1}{2}[h^{\mu\nu} - h^{\nu\mu}]$.  We have written the above action in euclidian space; the transition to lorentzian signature is straightforward.

 Performing the dimensional reduction starting from the action (\ref{BIact}) turns out to be rather cumbersome, as the lower-dimensional result can not immediately be rewritten in terms of $\Om^{i}{}_{\mu\nu}$. It turns out to be easier to first rewrite the action in a more convenient form and then reduce. When rewriting the action we ignore all terms which are proportional to the lowest order equation of motion
\be
0=\pa_{\mu}(\sqrt{h}h_{A}^{\mu\nu}) = -h_{S}^{\mu\eta}h_S^{\de\nu}\pa_{\mu}F_{\eta\de} \,,
\ee
since such terms can be removed by field redefinitions. We also remove all terms which vanish identically upon dimensional reduction when the gauge field in the lower dimension is set to zero. With these restrictions the action (\ref{BIact}) can be rewritten, after integrations by parts, as
\be \label{Invact}
-T_9\int \D^{10}x \, \sqrt{h}\,[1 - {\ts \frac{(2\pi\al')^2}{96}}(I_1 - 2 I_2  +  I_3 + 4 I_4) ]\,,
\ee
where
\bea \label{Invs}
&&\!\!\!\!\!\!\!\!I_1 = h_{S}^{\mu_1\mu_4}h_{S}^{\mu_2\mu_3}h_{S}^{\rho_1\rho_4}h_{S}^{\rho_2\rho_3} [\pa_{\rho_1}\pa_{\rho_2}F_{\mu_1\mu_2} + 2 \pa_{\rho_1}F_{[\mu_1|\nu_1} h_{A}^{\nu_1\nu_2} \pa_{\rho_2}F_{\mu_2]\nu_2} - \pa_{(\rho_1}F_{\rho_2)\nu_1}h_{A}^{\nu_1\nu_2}\pa_{\nu_2}F_{\mu_1\mu_2}] \non \\ && \qquad\qquad\qquad\qquad \;\, \times[\pa_{\rho_3}\pa_{\rho_4}F_{\mu_3\mu_4} + 2 \pa_{\rho_3}F_{[\mu_3|\nu_3} h_{A}^{\nu_3\nu_4} \pa_{\rho_4}F_{\mu_4]\nu_4} - \pa_{(\rho_3}F_{\rho_4)\nu_3}h_{A}^{\nu_3\nu_4}\pa_{\nu_4}F_{\mu_3\mu_4}] 
\non \\ && 
\!\!\!\!\!\!\!\!I_2 =  h_{S}^{\mu_1\mu_4}h_{S}^{\mu_2\mu_3}h_{S}^{\rho_1\rho_2}h_{S}^{\rho_3\rho_4} h_S^{\nu_1\nu_3} h_S^{\nu_2\nu_4} \pa_{\rho_1} F_{\mu_1\nu_1 }\pa_{\rho_2} F_{\mu_2\nu_2 }\pa_{\rho_3} F_{\mu_3\nu_3 }\pa_{\rho_4} F_{\mu_4\nu_4 } \,,
\non \\
&&\!\!\!\!\!\!\!\!I_3 =  h_{S}^{\mu_1\mu_2}h_{S}^{\mu_3\mu_4}h_{S}^{\rho_1\nu_3}h_{S}^{\rho_2\nu_4} h_{S}^{\nu_1\nu_2} h_S^{\rho_3\rho_4} \pa_{\rho_1} F_{\mu_1\nu_1 }\pa_{\rho_2} F_{\mu_2\nu_2 }\pa_{\rho_3} F_{\mu_3\nu_3 }\pa_{\rho_4} F_{\mu_4\nu_4 }\,,
\non \\
&&\!\!\!\!\!\!\!\!I_4 =  h_{S}^{\mu_1\mu_2}h_{S}^{\mu_3\rho_4}h_{S}^{\rho_3\mu_4}h_{S}^{\rho_1\rho_2} h_{S}^{\nu_1\nu_2} h_S^{\nu_3\nu_4} \pa_{\rho_1} F_{\mu_1\nu_1 }\pa_{\rho_2} F_{\mu_2\nu_2 }\pa_{\rho_3} F_{\mu_3\nu_3 }\pa_{\rho_4} F_{\mu_4\nu_4 } \,.
\eea
When dimensionally reducing, the gauge field $F_{MN}$ splits into $F_{\mu\nu}$ and $F_{\mu i} = \pa_{\mu}X_{i}$ ($F_{ij}\equiv 0$). Thus,
\be
h_{MN} = \left(\ba{cc} (1{+}F)_{\mu\nu} & \pa_{\mu}X_j \\ -\pa_{\nu}X_{i} & \de_{ij} \ea \right)\,.
\ee
The inverse, $h^{MN}$, becomes
\be
h^{MN} = \left(\ba{cc} \tilde{h}^{\mu\nu} & -\tilde{h}^{\mu\la}\pa_{\la}X^j \\ \pa_{\la}X^{i}\tilde{h}^{\la\nu} & \de^{ij} - \pa_{\la}X^{i}\tilde{h}^{\la\rho}\pa_{\rho}X^j \ea \right)\,,
\ee
where $\tilde{h}^{\mu\nu} = (\frac{1}{1 + F + \pa X^i \pa
X_i})^{\mu\nu}$. When $F_{\mu\nu}=0$, $\tilde{h}^{\mu\nu}$ reduces to
the inverse of the induced metric, $g^{\mu\nu}$.  Furthermore, it is important to note that (when $F$ is zero) $h^{ij} = P^{ij}$, where $P^{ij}$ is the projection operator defined below (\ref{Om}). 
 The dimensional reduction of $I_1$ leads to 
\bea \label{dimred1}
&& \!\!\!\!\!\!\!\!\!\!\!\!\!\!\!\!\!I_1 \rar  -2h^{ij}g^{\mu_1\mu_2}g^{\nu_1\nu_2}g^{\rho_1\rho_2} [\pa_{\mu_1}\pa_{\nu_1} \pa_{\rho_1} X_i - (\pa_{\mu_1}\pa_{\nu_1}X_{k}\pa_{\la}X^k g^{\la\eta}\pa_{\eta}\pa_{\rho_1}X_i + {\scriptstyle (\mu\rho\nu)} + {\scriptstyle (\rho\nu\mu) }   )] \non \\ && \qquad\quad\qquad\; \qquad \times  [\pa_{\mu_2}\pa_{\nu_2} \pa_{\rho_2} X_j - (\pa_{\mu_2}\pa_{\nu_2}X_{k}\pa_{\la}X^k g^{\la\eta}\pa_{\eta}\pa_{\rho_2}X_j + {\scriptstyle (\mu\rho\nu)} + {\scriptstyle (\rho\nu\mu) }   )]
\eea
which can be shown to be equal to $-2D_{\rho}\Om_{i\mu\nu} D^{\rho}\Om^{i\mu\nu}$ (where $D\Om$ is assumed to be projected onto the normal bundle). Similarly, the other $I_i$'s reduce as follows 
\bea \label{dimred2}
&& I_2 \rar 2\Om_{i\rho\mu}\Om^{i}{}_{\la\nu}\Om_{j}{}^{\rho\nu}\Om^{j\la\mu} \,, \non \\
&& I_3 \rar 2\Om_{i\rho\mu}\Om^{i\rho\la}\Om_{j\la\nu}\Om^{j\mu\nu} \,, \non \\
&& I_4 \rar\;\, \Om_{i\rho\mu}\Om^{i}{}_{\la\nu}\Om_{j}{}^{\rho\mu}\Om^{j\la\nu}  \,.
\eea
  In total we get the action (we have set $T_p=1$ and the $\cdot$ indicates contraction with $\de_{ij}$)
\bea \label{pdact1}
S &=& -\int \D ^{p+1}x \sqrt{g}\Big[ 1 + {\ts \frac{\pi^2\al'^2}{12} }(D_{\eta}\Om_{\mu\nu}\cdot D^{\eta}\Om^{\mu\nu} + 2 \Om_{\rho\mu}\cdot\Om_{\si\nu}\,\Om^{\rho\si}\cdot\Om^{\mu\nu} \non\\ &&\qquad\qquad \qquad\;-\; 2\Om_{\rho\mu}\cdot\Om_{\si\nu}\,\Om^{\rho\mu}\cdot\Om^{\si\nu} - \Om_{\rho\la}\cdot\Om^{\la\nu}\,\Om_{\nu\eta}\cdot\Om^{\eta\rho}) \Big]
\eea
which is written solely in terms of the second fundamental form as expected.

We would like to compare this action to the corrections obtained in \cite{Bachas:1999} (see also \cite{Fotopoulos:2001}). In that paper it was argued that the terms involving the second fundamental form should take the form 
\be \label{R2act}
S = - \int \sqrt{g}\Big(1 - {\ts \frac{(\pi\al')^2}{48} }[ (R_T)_{\rho\la\mu\nu}(R_T)^{\rho\la\mu\nu} - 2 (R_T)_{\mu\nu}(R_T)^{\mu\nu} - (R_N)_{\rho\la ij}(R_N)^{\rho\la ij} + 2\bar{R}_{ij}\bar{R}^{ij}]\Big)
\ee
where, in a flat ten-dimensional background, we have the relations \cite{Bachas:1999}
\bea \label{Riems}
(R_T)_{\mu\nu\rho\la} &=& \de_{ij}\Om^i_{\mu\rho}\Om^{j}_{\nu\la} - {\scriptstyle (\mu\leftrightarrow \nu)}\,, \non \\
(R_N)_{\mu\nu}{}^{ij} &=& g^{\rho\la}\Om^i_{\mu\rho}\Om^{j}_{\nu\la} - {\scriptstyle (\mu\leftrightarrow \nu) } \,,\non \\
\bar{R}^{ij} &=& g^{\mu_1\mu_2}g^{\nu_1\nu_2} \Om^i_{\mu_1\nu_1}\Om^j_{\mu_2\nu_2}\,.
\eea
Plugging these results into (\ref{R2act}) and dropping terms proportional to the lowest order equation of motion, $g^{\mu\nu}\Om^{i}{}_{\mu\nu}$ = 0, we get 
\be \label{pdact2}
S = -\int \D^{p+1}x \sqrt{g} [ 1 -{\ts \frac{\pi^2\al'^2}{12} }(\Om_{\rho\mu}\cdot\Om_{\si\nu}\,\Om^{\rho\mu}\cdot\Om^{\si\nu} - \Om_{\rho\mu}\cdot\Om^{\mu\nu}\,\Om_{\la\eta}\cdot\Om^{\eta\rho}) ]\,,
\ee
which at first sight differs from our result (\ref{pdact1}). However, it is possible to show that (when the background ten-dimensional space is flat) the following relation holds (modulo field redefinitions and total derivatives)\footnote{We would like to thank Angelos Fotopoulos for discussions which were helpful in establishing this relation.}
\be
 D_{\rho} \Om_{\mu\nu}\cdot D^{\rho}\Om^{\mu\nu} = 2\, \Om_{\rho\mu}\cdot \Om^{\mu\la}\,\Om_{\la\eta}\cdot \Om^{\eta\rho} + 
\Om_{\rho\mu}\cdot \Om_{\eta\de}\,\Om^{\rho\mu}\cdot \Om^{\eta\de} -2\, \Om_{\rho\mu}\cdot \Om_{\eta\de}\,\Om^{\rho\eta}\cdot \Om^{\mu\de}\,,
\ee
which is exactly what is needed for agreement between (\ref{pdact1}) and (\ref{pdact2}). We thus find agreement between our result and the result in \cite{Bachas:1999}. The result in \cite{Bachas:1999} was obtained by extrapolation from four-string scattering amplitudes, and thus the extension to all orders may not be unambiguous. 
 In fact, it was recently noted \cite{Fotopoulos:2001} that the result in \cite{Bachas:1999}, at least at first sight, does have some ambiguities, but it was argued that such ambiguities should be absent. Our result confirms that the result in \cite{Bachas:1999} is correct to all orders. The agreement between (\ref{pdact1}) and (\ref{pdact2}) is also indirect evidence for the correctness of the result (\ref{BIact}).

\setcounter{equation}{0}
\section{Derivative corrections to the Born-Infeld action and the Seiberg-Witten map} \label{SWsec}
In this section we will discuss the transformation properties of the derivative corrections to the D-brane action under the Seiberg-Witten (SW) map relating the ordinary and non-commutative versions of the D-brane effective action.

As is well known, the SW map is such 
that the D-brane effective action is form invariant after transforming the gauge field, the metric and the string coupling constant in prescribed ways, replacing ordinary products with $*$-products and ordinary derivatives with covariant derivatives. 
More precisely, the relations between the metrics and between the coupling constants are encoded in\footnote{We have rescaled $B$ by a factor of $2\pi\al'$; $B$ in the following formul\ae{} is really $2\pi\al 'B$. A similar comment holds for $\Phi$ and $\tha$. }: 

\be
\frac{1}{g+B} = \frac{1}{G+\Phi} + \tha \,, \qquad \frac{G_{\mathrm{S}}}{g_\mathrm{S}} = \left(\frac{\det{(G+\Phi)} } {\det{(g+B)} }\right)^{\frac{1}{2}}\,.
\ee
 The $*$-product is defined in the usual way as  
\be 
X(x)*Y(x) = e^{i\pi\al'\tha^{\eta\de}\pa^1_{\eta}\pa^2_{\de}}X(x^1)  Y(x^2)|_{x^1=x^2=x} = X Y  + i\pi\al'\tha^{\eta\de}\pa_{\eta}X\pa_{\de}Y + \cO(\al'^2) \,,
\ee 
and the covariant derivative is defined as $D_{\rho} X = \pa_{\rho}X  -  \frac{i}{2\pi\al'}[A_{\rho}*X - X* A_{\rho}]$. There are various different possible choices for $\Phi$ \cite{Seiberg:1999,Seiberg:2000}. Throughout this paper we will set $\Phi=0$ and work to first order in the non-commutativity parameter, $\tha$, but to all orders in $F$. 
In the recent work \cite{Das:2001} the transformation properties of the derivative corrections to the D-brane action in the superstring theory were discussed (both for the Born-Infeld and Wess-Zumino parts of the action) using the all order result for the SW map \cite{Liu:2000}, but working order by order in $F$. In that work the choice $\Phi=-B$ was made. Our approach is complementary to the one in \cite{Das:2001}.

\subsection{The zeroth-order terms}

Before proceeding to the derivative corrections, let us review the situation for the zeroth-order terms in the derivative expansion. 

\medskip
{\em The gauge field}
\medskip

As already mentioned the SW map transforms the lagrangean written in terms of the non-commutative variables $\cL(\hat{F},G,G_{\mathrm{S}},D,*)$ to the equivalent description in terms of the commutative variables and the lagrangean $\cL(F+B,g,g_{\mathrm{S}},\pa,\cdot)$. For simplicity we will expand all quantities in $B$ and work to first order in $B$. 
As we will see later, from an algebraic point of view this is essentially equivalent to working to first order in $\tha$. To first order in $B$ the transformation of the gauge field is  $\hat{F}=F+\de F$, with  
\be \label{deF}
\de F_{\mu\nu} = F_{\mu\eta}B^{\eta\de}F_{\de\nu} + A_{\eta} B^{\eta\de}\pa_{\de} F_{\mu\nu}\,.
\ee
The metric and the string coupling constant are unchanged at this
order. Neglecting all derivatives acting on $F$ implies that the $*$-products can be replaced by ordinary products in the action for the non-commutative theory. 
To the order we are working, the statement that $\cL(\hat{F})$ and $\cL(F+B)$ should be related by the SW map can be formulated as (modulo a total derivative)  $\De \cL(F) = 0$, where $\De = \de - \de_B$, with $\de$ is as in (\ref{deF}) and $\de_{B} F = B$. 
The equality of the non-commutative and ordinary actions can thus be formulated as an invariance of the action for the non-commutative theory under $\De$. By a slight abuse of terminology, we will refer to $\De = \de - \de_B$ as the SW map. In what follows, we will drop the hats from the non-commutative variables. One can shown that invariance under the SW map of the lagrangean $\cL(F)$ depending only of the gauge field strength, $F$, implies the differential equation  
\be
\frac{\de \cL}{\de F_{\mu\nu}} + \frac{\de \cL}{\de F_{\eta\de}} {F_{\eta}}^{\mu} {F_{\de}}^{\nu} - F^{\mu\nu}\cL = 0 \,.
\ee
By going to the special Lorentz frame where $F_{\mu\nu}$ is
block-diagonal with blocks 
$\left(\ba{cc} 0 & f_i \\ -f_i &0 \ea \right)$ it is easy to show that
(up to a multiplicative constant) 
the general solution is the Born-Infeld lagrangean $\sqrt{\det(1+F)}$. Thus, the SW map uniquely determines 
the Born-Infeld action as the only invariant action (if one neglects higher-derivative
terms)\footnote{Throughout this paper we will work in euclidian space. In the lorentzian case, the time direction should be treated more carefully.}. 

\medskip
{\em The transverse scalars}
\medskip

So far we have only discussed the dependence of the lagrangean on the gauge field. In dimensions less than ten, the D-brane action also depends on the transverse scalars. 
 We denote the transverse scalars by $X^i$, where $i=p{+}1,\ldots,9$. It is easy to derive the SW transformation $\De X^i \equiv \de X^i = -\tha^{\eta\de}A_\eta\pa_\de X^i$. The natural naive guess is that the lagrangean is a function of\footnote{We have rescaled the $X^i$'s to make $\pa X$ dimensionless.} $F_{\mu\nu}$ and $\pa_\mu X^i\pa_\nu X_i$: $\cL = \cL(F,\pa X*\pa X)$. Requiring this action to be invariant under $\De$ leads to the equation:
\be \label{Xde}
\frac{\de \cL}{\de F_{\mu\nu}} = -\frac{\de \cL}{\de F_{\eta\de}} F_\eta{}^\mu F_\de{}^\nu + F^{\mu\nu}\cL - 2 \pa_{\eta}X^i F^{\mu}{}_{\de}\pa^{\nu}X_i\frac{\de \cL}{\pa_\eta X \pa_\de X}\,.
\ee
It is easy to see that the expected result $\det(g+F+\pa X \pa X)^{-\frac{1}{2}}$ does not satisfy this relation. 
The reason for the discrepancy is that on the non-commutative side there can be commutator terms of the form $\frac{i}{2\pi\al'}[X^i,X^j] = B^{\eta\de}\pa_\eta X^i\pa_\de X^j + \cO(B^2)$, which lead to corrections to the map $\De$ and hence to the above equation (\ref{Xde}). To see which terms are required we recall that in \cite{Tseytlin:1997} the terms needed for consistency of the {\em non-abelian} D-brane action with T-duality were determined with the result 
\be
S = -T_p\int d^{p+1}\xi\, \Tr(\det D \det Q)^{\frac{1}{2}}\,,
\ee
where the matrices $D_{\mu\nu}$ and $Q_{\mu\nu}$ are given in \cite{Tseytlin:1997}. We will only need the leading terms here: $D_{\mu\nu} = (g + F)_{\mu\nu} - \frac{i}{2\pi\al'}\pa_{\mu}X^i[X^i,X^j]\pa_{\nu}X^j +\ldots$ and $\det Q = 1 + \ldots$. 
Reinterpreting this action as an action for the non-commutative U(1) theory (i.e.\ replacing matrix products with $*$-products and then taking the gauge group to be $\mathrm{U}(1)$) shows that the extra contribution coming from $\De \sqrt{\det D}$ (more precisely from the part with the $[X^i,X^j]$ commutator) is exactly what is needed to make the action invariant (to first order in $B$). 
Turning things around, one could have used the requirement of invariance under the SW map together with the known form of the action in the commutative limit to learn about ``commutator'' corrections in the non-commutative U(1) theory. These can be unambiguously lifted to corrections in the non-commutative U($N$) theory if one assumes that the action only contains an overall trace (as is expected) and can then be translated into corrections in the non-abelian U($N$) theory by taking the commutative limit. 

\medskip
{\em The fermions}
\medskip

One may also study how the introduction of fermions affects the discussion. The fermions transform as $\De\Psi \equiv \de \Psi = -\tha^{\eta\de}A_\eta\pa_\de\Psi$ under the SW map. The form of this transformation is fixed by the fact that the fermions in the $\mathrm{U}(1)$ commutative theory are invariant under gauge transformations. 
Let us assume that the action contains the usual kinetic term (for a fermion coupled to a gauge field) $\cL^{(0)} = \bar{\Psi}\ga^\la D_\la \Psi$. Proceeding naively, one finds that  $\De\cL^{(0)} = -\frac{1}{2}\tha^{\eta\de}F_{\eta\de}\bar{\Psi}\ga^\la\pa_\la\Psi + 
\tha^{\eta\de}F_{\eta\la}\bar{\Psi}\ga^\la\pa_\de\Psi$. Adding the most general term involving $F^2$: 
\be  
a F^{\mu\nu}F_{\mu\nu} \bar{\Psi} \ga^\la D_\la\Psi + b F^{\la \mu}F_{\mu}{}^\nu \bar{\Psi}\ga_\la D_\nu\Psi \,,
\ee
 and choosing $a=\frac{1}{4}$, $b=1$ implies that its variation cancels the variation of $\cL^{(0)}$ (modulo $F^3$ terms), however, it also gives an additional term: $B^{\eta\de}F_{\de}{}^{\la}\bar{\Psi}\ga_\eta\pa_\la\Psi$, 
which seemingly can not be removed. The resolution of this puzzle can again be traced to a ``non-abelian'' effect. 
Recall that $\ga^\la$ is the pull-back to the world volume of $\Ga^M$, the ten-dimensional gamma matrix, i.e.\ $\ga_{\la} = \pa_\la Y_M\Ga^M$. Now in the ordinary action we use the static gauge so $\pa_\rho Y^M = \de^M_\rho$ (we ignore the transverse scalars if present). In the non-commutative action we should replace the derivative appearing in the pull-back by the covariant derivative, $D_{\rho}$. 
This requirement was first obtained in the context of the non-abelian D-brane action in \cite{Hull:1998}. With this modification we find that $\ga_\la$ transforms as $\ga_\la \rar -B^{\eta\de}F_{\eta\la}\ga_\de$, which gives rise to precisely the missing term required for invariance. 

More generally,  one can assume a lagrangean of the form $M^{\mu\nu}(F)\bar{\Psi}\ga_\nu\pa_\nu\Psi$. Requiring invariance under the SW map leads to the equation
\be
(F_{\rho \eta}B^{\eta\de} F_{\de \la} - B_{\rho\la})\frac{\de M^{\mu\nu}}{\de F_{\rho\la}} + B^{\mu\eta}F_{\eta\de}M^{\de \nu} + M^{\mu\eta}F_{\eta\de}B^{\de \nu} + {\ts \frac{1}{2}}B^{\eta\de}F_{\eta\de}M^{\mu\nu}=0 \,,
\ee
which is solved by  (assuming that $M^{\mu\nu} \propto \de^{\mu\nu}+\ldots$ for small $F$): $M^{\mu\nu}\propto \sqrt{\det h}\, h_{S}^{\mu\nu}$. This can be shown by making a general ansatz of the form $M^{\mu\nu} = \sum_p g_p(F) (F^p)^{\mu\nu}$. 
By a suitable rescaling of the $\Psi$'s, we find that the above terms arise form the expansion of $\sqrt{\det( h_{\mu\nu} + \bar{\Psi}\ga_{(\mu}\pa_{\nu)}\Psi)}$ to second order in $\Psi$. We would like to compare this expression to the two-fermion part of the full (gauge-fixed) D-brane action \cite{Aganagic:1996b}. 
In this action, the argument of the determinant is $h_{\mu\nu} + \bar{\Psi} \ga_{(\mu} \pa_{\nu)} \Psi) + B_{\mu\nu}+\bar{\Psi}\ga_{[\mu}\pa_{\nu]}\Psi$. It thus appears that it is the combination $B_{\mu\nu}+\bar{\Psi}\ga_{[\mu}\pa_{\nu]}\Psi $ which should be identified with the $B$ appearing in the SW map. One may also investigate invariance of the full gauge-fixed $\ka$-symmetric action (which includes higher powers of the fermions), but we will not do so here. Note that the SW map does not mix terms with different numbers of fermions. To relate different orders in the expansion in powers of the fermions one needs to use another approach, such as for instance the one described in \cite{Cederwall:2001a}.

\subsection{Derivative corrections: general framework}

For the discussion of the derivative corrections it is convenient to introduce some further terminology (related discussions can be found in \cite{Terashima:2000,Cornalba:1999}; in particular, some of the results below were also obtained in \cite{Terashima:2000}). 

We will call an expression $T_{\mu_1,\ldots,\mu_k}$ a covariant tensor under the SW map (or simply a tensor when no confusion is likely to arise) if it transforms as: 
\be \label{covT}
T_{\mu_1\ldots,\mu_2} \rar F_{\mu_1\eta}B^{\eta\de}T_{\de\mu_2,\ldots,\mu_n} + \cdots + F_{\mu_n\eta}B^{\eta\de}T_{\mu_1\ldots\mu_{n-1}\de} + A_{\eta}B^{\eta\de}\pa_{\de}T_{\mu_1,\ldots,\mu_n}\,.
\ee
Similarly, a contravariant tensor under the SW map is a quantity which transforms as 
\be \label{conT}
T^{\mu_1\ldots\mu_n} \rar -B^{\mu_1\eta}F_{\eta\de}T^{\de\mu_2,\ldots,\mu_n} - \cdots - B^{\mu_n\eta}F_{\eta\de}T^{\mu_1\ldots\mu_{n-1}\de} + A_{\eta}B^{\eta\de}\pa_{\de}T^{\mu_1,\ldots,\mu_n}\,.
\ee
 Tensors with mixed index structure are 
defined in the usual way. It follows from a short calculation that any expression of the form $\frac{1}{g_s}\sqrt{h}S$, where $S$ is a scalar (i.e.\ formed out of fully contracted tensors of the above form) is invariant under the SW map to first order. 
Moreover, the scalar $S$ is invariant under the above transformations without the $A_{\eta}B^{\eta\de}\pa_\de$ part. An example of a covariant tensor is $D_\rho F_{\mu\nu}$; note that $F_{\mu\nu}$ itself is not a tensor, nor is $D_{\rho}D_{\la}F_{\mu\nu}$. 
An example of a contravariant tensor is $h_{S}^{\mu\nu}$. Indices can thus be raised using $h_S^{\mu\nu}$ (and lowered with its inverse, which is also a tensor since $\de^{\mu}_{\nu}$ is) without violating the tensor property, although we will not use this convention; all occurences of $h_{S}^{\mu\nu}$ are written explicitly. On the other hand $h_A^{\mu\nu}$ does not satisfy (\ref{covT}) and hence is not a tensor. It does however have a simple transformation rule 
\be
\De h_{A}^{\mu\nu} = B^{\mu\nu} - B^{\mu\eta}F_{\eta\de}h_{A}^{\de\nu} - B^{\nu\eta}F_{\eta\de}h_{A}^{\mu\de} + A_{\eta}B^{\eta\de}\pa_{\de}h_{A}^{\mu\nu}\,.
\ee 
We will also need the transformation of $D_{\rho}D_{\la}F_{\mu\nu}$:
\be
\De D_{\rho}D_{\la}F_{\mu\nu} = B^{\eta\de}[\pa_{\rho}\pa_{\la}(F_{\mu\eta}F_{\de\nu}) + \pa_{\rho}(F_{\la\eta}\pa_{\de}F_{\mu\nu}) - F_{\rho\eta}\pa_{\de}\pa_{\la}F_{\mu\nu}] + A_{\eta}B^{\eta\de}\pa_{\de}\pa_{\rho}\pa_{\la}F_{\mu\nu}\,.
\ee
Comparing the above formul\ae{} to the related ones appearing in \cite{Terashima:2000} (where the restriction to first order in $B$ was not used), we see that after replacing $B$ by $-\tha$ they  become equal; thus, from an algebraic point of view restricting to small $B$ is actually not a limitation. 
Using the  above results one can show that although $D_{\rho}D_{\la}F_{\mu\nu}$ is not a tensor, the following expression is:
\be \label{DDFinv}
D_{\rho}D_{\la}F_{\mu\nu} - D_\rho F_{\la\eta}h_{A}^{\eta\de}D_{\de}F_{\mu\nu} - D_{\rho}F_{\mu \eta}h_A^{\eta\de}D_{\la}F_{\de\nu} - D_{\rho}F_{\nu \eta}h_A^{\eta\de}D_{\la}F_{\mu\de}\,.
\ee
This suggests the following definition of a covariant derivative acting on a covariant tensor 
\be \label{covD}
\cD_\rho T_{\mu_1,\ldots,\mu_k} = D_{\rho}T_{\mu_1,\ldots,\mu_k} - D_\rho F_{\mu_1\eta}h_{A}^{\eta\de}T_{\de,\ldots,\mu_k} -\cdots - D_{\rho}F_{\mu_k\eta}h_{A}^{\eta\nu}T_{\mu_1,\ldots,\mu_k}\,.
\ee
It is straightforward to check that this expression indeed transforms as a covariant tensor. Similarly, acting on a contravariant tensor
\be \label{conD}
\cD_\rho T^{\mu_1,\ldots,\mu_k} = D_{\rho}T^{\mu_1,\ldots,\mu_k} + h_A^{\mu_1\eta} 
D_\rho F_{\eta\de} T^{\de,\ldots,\mu_k} +\cdots +h_A^{\mu_k\eta} D_{\rho}F_{\eta\de}T^{\mu_1,\ldots,\mu_k} \,,
\ee
one obtains a mixed tensor.  Notice that with this definition $\cD_{\rho}h_{S}^{\mu\nu} = 0$. Computing the commutator of two $\cD$'s one finds (to zeroth order in the non-commutativity parameter)
\be \label{defrel}
{}[\cD_{\rho},\cD_{\la}]V_{\mu} = -[\pa_{\rho}F_{\mu\eta}h_{S}^{\eta\de}\pa_{\la}F_{\de\nu}h_{S}^{\nu\ga} - {\scriptstyle (\rho{\leftrightarrow}\la)}]V_{\ga} + \pa_{\eta}F_{\rho\la}h_{A}^{\eta\ga}\cD_{\ga}V_{\mu}\,.
\ee
By comparing this expression to the usual result ${}[\cD_{\rho},\cD_{\la}]V_{\mu} = -R^{\ga}{}_{\mu\rho\la}V_{\ga} + T^{\ga}{}_{\rho\la}\cD_{\ga}V_{\mu}$, one can read off the Riemann tensor and the torsion: 
\be \label{Rietor}
R^{\ga}{}_{\mu\rho\la}= \pa_{\rho}F_{\mu\eta}h_{S}^{\eta\de}\pa_{\la}F_{\de\nu}h_{S}^{\nu\ga} - {\scriptstyle (\rho{\leftrightarrow}\la)}\,, \qquad  T^{\ga}{}_{\rho\la} = -h_{A}^{\ga\eta}\pa_{\eta}F_{\rho\la}\,.
\ee
There is an apparent contradiction since the torsion is not a tensor under the SW map; this is an artefact of taking the commutative limit. Retaining the $D$'s one gets an additional term on the right-hand side of the defining relation (\ref{defrel}): $[D_{\rho},D_{\la}]V_{\mu} = -\frac{i}{2\pi\al'}[F_{\rho\la},V_{\mu}]$. This term compensates the unwanted part of the transformation of $T^{\ga}{}_{\rho\la}\cD_{\ga}V_{\mu}$. 

We will end this section with a few remarks. Above we used the ``canonical'' SW map. In general, field redefinitions will change the SW map. It is easy to see that a tensor will generically not remain a tensor after a field redefinition. The ambiguity of the SW map and the relation to field redefinitions was first pointed out in \cite{Asakawa:1999} (see also \cite{Okawa:2000}). The possibility of derivative corrections to the SW map will be discussed later on.

\subsection{Bosonic-string D-branes: two derivative corrections}

Before  discussing the D-branes in the superstring theory we will briefly discuss the case of the two-derivative corrections to the action for a D-brane in the bosonic string theory. The two-derivative corrections involving four powers of the gauge field strength, $F$, were calculated several years ago \cite{Andreev:1988}. 
More, recently, an extension of this result to all orders in $F$ was proposed using the partition function approach with a proper treatment of the tachyon \cite{Andreev:2001,Tseytlin:2000}. We will now rederive the result in \cite{Andreev:2001} via a different more heuristic route. Up to field redefinitions the fourth-order expression for the two-derivative corrections is given by (to indicate the relative normalisation, we have also included the zeroth-order term)
\be \label{4dJ}
S = - \int \sqrt{h} + \frac{2\pi\al'}{48\pi}(4 J_3 - 8 J_2 - J_1)\,,
\ee
where 
\be J_1 = F_{\eta\de}F^{\de\eta} \pa_{\rho}F_{\mu\nu}\pa^{\rho}F^{\nu\mu}\,, \quad J_2 = F^{\rho\de}F_{\de}{}^{\la}\pa_{\eta}F_{\rho\mu}\pa^{\eta}F^{\mu}{}_{\la}\,, \quad J_3 = F^{\rho\de}F_{\de}{}^{\la}\pa_{\rho}F_{\mu\eta}\pa_{\la}F^{\eta\mu} \,.
\ee 
 With the assumption that the all-order result for the two-derivative corrections should be expressible in terms of $h^{\mu\nu}$ and $\pa^k F$'s only, a basis of all possible two-derivative terms (modulo field redefinitions and integrations by parts) is
\bea
X_1 &=& \sqrt{h}\, h_{A}^{\mu\nu}h_{A}^{\eta\de}h_{S}^{\rho\la}\pa_{\rho}F_{\mu\nu}\pa_{\la}F_{\eta\de} \,, \non \\
X_2 &=& \sqrt{h}\, h_{A}^{\mu\nu}h_{A}^{\eta\de}h_{S}^{\rho\la}\pa_{\mu}F_{\eta\rho}\pa_{\de}F_{\nu\la} \,, \non \\
X_3 &=& \sqrt{h}\, h_{S}^{\mu\nu}h_{S}^{\eta\de}h_{S}^{\rho\la}\pa_{\rho}F_{\mu\eta}\pa_{\la}F_{\nu\de} \,, \non \\
X_4 &=& \sqrt{h}\, h_{A}^{\mu\nu}h_{A}^{\eta\de}h_{S}^{\rho\la}\pa_{\mu}F_{\eta\rho}\pa_{\nu}F_{\de\la} \,.
\eea
For completeness we note that
\be
X_5 = \sqrt{h}\, h_{A}^{\mu\nu}h_{A}^{\rho\la}h_{S}^{\eta\de}\pa_{\eta}F_{\de\rho}\pa_{\la}F_{\mu\nu}\,, \qquad X_6 = \sqrt{h}\, h_{A}^{\mu\nu}h_{S}^{\rho\la}\pa_{\rho}\pa_{\la}F_{\mu\nu} \,,
\ee 
can be related to the above basis elements via the identities (total derivatives as well as terms removable by field-redefinitions are treated as 0)
\bea
0  &=& \sqrt{h}\, h_A^{\mu\nu}\pa_{\mu}(h_S^{\rho\la}\pa_{\rho}F_{\la\nu}) = {\ts \frac{1}{2}}X_6-2X_2 + X_4 \,,\non \\
0 &=& \pa_{\rho} (\sqrt{h} \,h_A^{\mu\nu}h_S^{\rho\la} \pa_{\la} F_{\mu\nu}) = X_6 - 2X_2 - X_3 +2X_4 - X_5\,.
\eea
The most general action consistent with the requirement that it reduces to (\ref{4dJ}) when restricting to the fourth-order terms is
\be \label{BallF}
S = -\int \sqrt{h}\bigg( 1 + {\ts \frac{2\pi\al'}{12\pi} }[ 4 X_4 + \ga X_3 -2(3{-}\ga)X_2] \bigg) \,,
\ee
where $\ga$ is undetermined by our approach. To determine $\ga$ one would need to use another approach, such as the $\beta$-function method \cite{Andreev:1988b}. The above result agrees with the more careful derivation given in \cite{Andreev:2001}. 

Before proceeding to discuss the transformation properties of the above action under the SW map, let us discuss the dimensional reduction of the above action. Using the methods of section \ref{sLow} and retaining only the transverse scalars in the lower dimension we find 
\bea
X_2 &\rar& - g^{\mu\la}\pa_{\la}X^ig^{\mu\de}\pa_{\de}X^j\pa_{\mu}\pa_{\rho}X_j \pa_{\nu}\pa_{\la}X_i \,,\non \\
X_3 &\rar& 2h^{ij}g^{\rho\la}g^{\mu\nu}\pa_{\rho}\pa_{\mu}X_i\pa_{\la}\pa_{\nu}X_j = 2 \Om_{\rho\mu}\cdot \Om^{\rho\mu} \,,\non \\
X_4 &\rar& 0\,.
\eea
We see that $X_3$, but not $X_2$, can be written in terms of the second fundamental form, $\Om^{i}{}_{\mu\nu}$. This seems to suggest that $X_2$ should not appear in the action, which would imply  $\ga=3$. Furthermore, the numerical coefficient in front of the $\Om^2$ term then agrees with the recent result \cite{Corley:2001}. We would also like to point out that only for $\ga=3$ is it possible to write the action (\ref{BallF}) in terms of $h^{\rho\la}$ and $S_{\rho\la\mu\nu}$ (\ref{S}) only. This can e.g.\ be accomplished by using $X_3 = 2 h_S^{\rho\la}h_S^{\mu\nu}S_{\rho\mu\nu\la}$ and $X_4 = \frac{1}{2}h_A^{\rho\la}h_A^{\mu\nu}S_{\rho\la\mu\nu}$.

We now turn to the transformation properties of the above action under the SW map. From the general discussion about invariants presented earlier it is easy to see that there is only one two-derivative invariant that can be constructed, namely $X_3$. Thus, the action (\ref{BallF}) appears not to be invariant under the SW map. 
This was first observed in \cite{Okawa:1999} for the fourth-order terms.  A resolution was later proposed \cite{Okawa:2000} which resolved the problem by modifying the SW map at order $\al'$. We will now check if a similar procedure also works to all orders in the gauge field. First we note that
\be \label{DX4}
\De X_4 = \sqrt{h}\, B^{\mu\nu}h_{A}^{\eta\de}h_{S}^{\rho\la}\pa_{\mu}F_{\eta\rho}\pa_{\nu}F_{\de\la} +  \sqrt{h}\, h_{A}^{\mu\nu} B^{\eta\de}h_{S}^{\rho\la}\pa_{\mu}F_{\eta\rho}\pa_{\nu}F_{\de\la}\,.
\ee
The first term can be canceled by adding the term $\frac{i}{3\pi}\sqrt{h}h_{A}^{\eta\de}h_{S}^{\rho\la} [F_{\eta\rho},F_{\de\la}]$ to the lagrangean of the non-commutative theory. 
Reinterpreting this correction as a correction to the non-abelian action: $\frac{i}{3\pi}\Tr(\sqrt{h}h_{A}^{\eta\de}h_{S}^{\rho\la} [F_{\eta\rho},F_{\de\la}]) = \frac{i}{3\pi}\Tr(F^{\eta\de}[F_{\eta\rho},F_{\de}{}^{\la}]) + \ldots$, one finds that the coefficient of the $F^3$ term agrees with the known result \cite{Tseytlin:1986} for the non-abelian action (after taking into account differences in conventions). Thus, the first term in (\ref{DX4}) can be understood, but what about the second term? 
In order for a term of the form $\sqrt{h}h_A^{\mu\nu} V_{\mu\nu}$ to be removable by a modification of the SW map for the gauge field, $A_\mu$ it is necessary that $\pa_{[\rho}V_{\mu\nu]}=0$. 
The second term in (\ref{DX4}) corresponds to $V_{\mu\nu} = B^{\eta\de}h_S^{\rho\la}\pa_{\mu}F_{\eta\rho}\pa_{\nu}F_{\de\la}$, which does not satisfy the requirement $\pa_{[\rho}V_{\mu\nu]}=0$ and thus can not be removed by modifying the SW map. We note that the leading order term $V_{\mu\nu} = B^{\eta\de}\de^{\rho\la}\pa_{\mu}F_{\eta\rho}\pa_{\nu}F_{\de\la}$ does, however, satisfy the requirement  $\pa_{[\rho}V_{\mu\nu]}=0$ and can thus be removed;  
this is the modification discussed in \cite{Okawa:2000}. We conclude that if the action (\ref{BallF}) is the correct extension to all orders in $F$ then it appears not to be invariant under the SW map, even if one allows for modifications of the form discussed in \cite{Okawa:2000}. 
A possible resolution might be that the tachyon should have a non-trivial $F$-dependent transformation under the SW map.

\subsection{Superstring D-branes: four-derivative corrections}

We will now discuss the four-derivative corrections \cite{Wyllard:2000} to the  parity-conserving part of the D9-brane action in type II superstring theory (cf.\ (\ref{BIact})). 
However, before turning to the four-derivative corrections, let us discuss the two-derivative corrections. It is known that the two-derivative corrections to the action for the commutative theory vanish. In the action for the non-commutative theory there might be additional commutator terms of the form $K^{\mu\nu\eta\de}(F)[F_{\mu\nu},F_{\eta\de}]$. 
The $\cO(B)$ part of such a commutator term could be canceled (assuming $K^{\mu\nu\eta\de}(F)$ is a tensor) by the term $2\pi\al'h_A^{\rho\la}K^{\mu\nu\eta\de}(F)D_{\rho}F_{\mu\nu}D_{\la}F_{\eta\de}$, but since such a term survives in the commutative limit it has to be absent.  
Thus one can argue that the absense of the two-derivative corrections in the commutative action implies the absense of ``one-commutator'' corrections in the non-commutative action. 
We have not really proven this statement here, since one would also need to show that the there are no commutator terms which transform into terms which are either total derivatives or can be removed by field redefinitions. 

Let us now turn to the four-derivative corrections to the D9-brane action (\ref{BIact}). By inspection it is easy to see that the action is not related in any simple way to tensors under the SW map, cf.\ the discussion earlier in this paper. It turns out to be very involved to rewrite the action in terms of invariants by using integrations by parts etc. We will therefore only discuss a subset of the corrections here, namely the terms which vanish upon dimensional reduction when setting the gauge field in the lower dimension to zero. 
These terms were identified in section \ref{sLow}, where the action was also written in a form convenient for performing the dimensional reduction (\ref{Invact}), (\ref{Invs}). It turns out that this way of writing the corrections is also the most convenient form to use when discussing the invariance under the SW map.
 
Using earlier results it is easy to see that the action (\ref{Invact}) can be rewritten as 

\be \label{Invact2}
-T_9\int \D^{10}x \, \sqrt{h}\,[1 - {\ts \frac{(2\pi\al')^2}{96}}(I_1 - 2 I_2  +  I_3 + 4 I_4) ]\,,
\ee
where
\bea \label{Invs2}
&&\!\!\!\!\!I_1 = h_{S}^{\mu_1\mu_4}h_{S}^{\mu_2\mu_3}h_{S}^{\rho_1\rho_4}h_{S}^{\rho_2\rho_3} \cD_{(\rho_1}D_{\rho_2)}F_{\mu_1\mu_2} \cD_{(\rho_3}D_{\rho_4)}F_{\mu_3\mu_4}\,,
\non \\ && 
\!\!\!\!\!I_2 =  h_{S}^{\mu_1\mu_4}h_{S}^{\mu_2\mu_3}h_{S}^{\rho_1\rho_2}h_{S}^{\rho_3\rho_4} h_S^{\nu_1\nu_3} h_S^{\nu_2\nu_4} D_{\rho_1} F_{\mu_1\nu_1 }D_{\rho_2} F_{\mu_2\nu_2 }D_{\rho_3} F_{\mu_3\nu_3 }D_{\rho_4} F_{\mu_4\nu_4 } \,,
\non \\
&&\!\!\!\!\!I_3 =  h_{S}^{\mu_1\mu_2}h_{S}^{\mu_3\mu_4}h_{S}^{\rho_1\nu_3}h_{S}^{\rho_2\nu_4} h_{S}^{\nu_1\nu_2} h_S^{\rho_3\rho_4} D_{\rho_1} F_{\mu_1\nu_1 }D_{\rho_2} F_{\mu_2\nu_2 }D_{\rho_3} F_{\mu_3\nu_3 }D_{\rho_4} F_{\mu_4\nu_4 }\,,
\non \\
&&\!\!\!\!\!I_4 =  h_{S}^{\mu_1\mu_2}h_{S}^{\mu_3\rho_4}h_{S}^{\rho_3\mu_4}h_{S}^{\rho_1\rho_2} h_{S}^{\nu_1\nu_2} h_S^{\nu_3\nu_4} D_{\rho_1} F_{\mu_1\nu_1 }D_{\rho_2} F_{\mu_2\nu_2 }D_{\rho_3} F_{\mu_3\nu_3 }D_{\rho_4} F_{\mu_4\nu_4 }  \,.
\eea
We conclude that all the $I_i$'s are built out of tensors and thus the action (\ref{Invact2}) is manifesly invariant. To complete the proof of the invariance of the four-derivative corrections (\ref{BIact}) under the SW map, one has to show that the remaining terms can also be written in terms of invariants or that their variation can be canceled by adding commutator terms to the action or that they can be removed by field redefinitions. We hope to return to this question elsewhere. 

We will end this section with a discussion of the various Riemann tensors and covariant derivatives that have appeared in this paper. Firstly, we have the Riemann tensor (\ref{Rietor}) associated with the covariant derivative (\ref{covD}), (\ref{conD}) and the SW map (\ref{covT}), (\ref{conT}). 
Secondly, we have the Riemann tensors (\ref{Riems}) associated with the embedding of a lower-dimensional D-brane into ten-dimensional space. The tensors and invariants of these two classes appear to be related. For instance, the invariants under the SW map (cf.\ (\ref{Invs2})) descend into invariants (cf.\ (\ref{dimred1}), (\ref{dimred2}) ) expressed in terms of the second fundamental form. Notice also the similarlty between the two Riemann tensors (\ref{Riems}) and (\ref{Rietor}). 

Yet another Riemann tensor also appeared above, namely the one (\ref{S}) in terms of which the derivative corrections (\ref{BIact}) are expressed. This Riemann tensor is constructed from the non-symmetric metric $h_{\mu\nu}$ and has an associated covariant derivative, $D_{\rho}$, defined as (note the placement of the indices on $\Ga^{\la}{}_{\mu\rho}$): 
\be
D_{\rho}T_{\mu_1,\ldots,\mu_k} = \pa_{\rho} T_{\mu_1,\ldots,\mu_k} - \Ga^\la{}_{\mu_1\rho}T_{\la,\ldots,\mu_k} - \ldots - 
\Ga^\la{}_{\mu_k\rho}T_{\mu_1,\ldots,\la}\,,
\ee
with $\Ga^\la{}_{\mu\rho} = h^{\la\si}\pa_{\mu}F_{\si\rho}$. Computing the commutator of two covariant derivatives we get $[D_{\mu},D_{\nu}] = -S^{\la}{}_{\si\mu\nu}V_{\la} + T^\la{}_{\mu\nu}D_{\la}V_{\si}$, with $S^{\la}{}_{\si\mu\nu} = h^{\la\de}S_{\de\si\mu\nu}$, 
where $S_{\de\si\mu\nu}$ is as in (\ref{S}), and the torsion $T^\la{}_{\mu\nu}$ is given by $T^\la{}_{\mu\nu} = h^{\la\si}\pa_{\si}F_{\mu\nu}$. Let us try to relate these results to the covariant derivative and Riemann tensor associated with the SW map. 
The connection associated to the covariant derivative appearing in (\ref{covT}) is $\Ga^{\la}{}_{\mu\rho} = h_A^{\la\si}\pa_{\rho}F_{\si\mu}$. This expression involves $h_A^{\mu\nu}$ rather than $h^{\mu\nu}$, but since $h_S^{\la\si}\pa_{\rho}F_{\si\mu}$ is a tensor we may always add such a term to the connection, which gives $\Ga^{\la}{}_{\mu\rho} = h^{\la\si}\pa_{\rho}F_{\si\mu}$. Comparing this connection to the one associated with $S_{\rho\la\mu\nu}$ we see that the two connections have the same symmetric part $\Ga^{\la}{}_{(\rho \mu)}$, but the antisymmetric parts (and the hence the torsions) are opposite in sign. The associated Riemann tensors are rather different since one has a $\pa^2 F$ part (\ref{S}), whereas the other does not (adding the above tensor to the connection does not change this aspect of (\ref{Rietor})). The interplay and relations between the different Riemann tensors might merit further investigation. 

Let us also remark that the results obtained in the recent work \cite{Das:2001}, where the choice $\Phi=-B$ was used, indicate that with this choice the natural framework in which to address the invariance of the derivative corrections is the one in which the corrections are written in terms of $S_{\rho\la\nu\mu}$. The fact that for our choice of $\Phi$ ($\Phi=0$) another connection was more natural suggests that there might be some relation between the choice of connection and the choice of $\Phi$.

\setcounter{equation}{0}
\section{Speculations on higher-order corrections}  \label{sSpec}

In this paper we have concentrated on four-derivative corrections. An interesting question is to what extent it is possible to obtain information about the corrections at higher orders in the derivative expansion. Explicit calculations become very involved, and are not a practical option (at least for the Born-Infeld part; the calculations for the Wess-Zumino term are somewhat simpler and certain higher order corrections were determined  in \cite{Wyllard:2000}). One possible approach would be to use the SW map and the methods discussed in section \ref{SWsec} to try to constrain the corrections. Unfortunately the number of possible invariants grows very quickly with the number of derivatives, and there is no simple method to fix the undetermined coefficients (see, however, \cite{Das:2001} for a more promising approach). Because of these difficulties we will try another more speculative approach. Our starting point is the following relation:
\bea \label{amusing}
&&-T_9\,e^{ \frac{\zeta(2)}{2}(\al'^2) \tr(\mathsf{S}_{\mu_1\mu_2}\mathsf{S}_{\mu_3\mu_4})\frac{\de}{\de F_{\mu_2\mu_1}}\frac{\de}{\de F_{\mu_4\mu_3}} } \sqrt{ h } = \non \\ &&
-T_9\left( 1 + {\ts \frac{(2\pi\al')^2}{96}}\bigg[\!\! -\!h^{\mu_4\mu_1}h^{\mu_2\mu_3}h^{\rho_4\rho_1}h^{\rho_2\rho_3}S_{\rho_1\rho_2\mu_1\mu_2}S_{\rho_3\rho_4\mu_3\mu_4}
  + \half
h^{\rho_4\rho_1}h^{\rho_2\rho_3}S_{\rho_1\rho_2}S_{\rho_3\rho_4}\bigg] \right) \sqrt{ h } \non \\&&+\cO(\al'^4)\,,
 \eea
where we have used the definition $(\mathsf{S}^{\rho_1}{}_{\rho_2})_{\mu_1\mu_2} := h^{\rho_1\la}S_{\la\rho_2\mu_1\mu_2}$ and $\tr$ denotes the ordinary trace over the $\rho$ indices.  
We note that the right-hand side of the above relation is precisely equal to the expression for the four-derivative corrections in eq.\ (\ref{BIact}).  
The left-hand side of the above relation (\ref{amusing}) is reminiscent of the expression for the derivative corrections to the Wess-Zumino term determined in \cite{Wyllard:2000}
\be \label{WZact}
S_{\mathrm{WZ}} = T_9 \int \exp [ \sum_{n\ge2} \ts{\frac{\ze(n)}{n}} \tr(\al'\mathsf{S})^{n} ] C e^F\,,
\ee
where we have suppressed all wedge products and $\mathsf{S}$ is the matrix-valued two-form $\mathsf{S}^{\rho_1}{}_{\rho_2} := \ts{\frac{1}{2!}}h^{\rho_1\la}S_{\la\rho_2\mu_1\mu_2}\D x^{\mu_1}\we \D x^{\mu_2}$.  
Notice that we have raised one index using $h^{\rho_1\la}$ compared to the definition used in \cite{Wyllard:2000} (this change allows us to use an ordinary trace (denoted by $\tr$) instead of the construction used there). 
The first term in the sum in the exponent in (\ref{WZact}), $\ts{\frac{\ze(2)}{2}} \tr(\al'\mathsf{S})^{2}$, becomes equal to the exponent in the expression on the left hand side of the equality (\ref{amusing}) if one makes the substitution $\tr(\mathsf{S}\we\mathsf{S}) \rar\tr(\mathsf{S}_{\mu_2\mu_1}\mathsf{S}_{\mu_4\mu_3})\frac{\de}{\de F_{\mu_1\mu_2}}\frac{\de}{\de F_{\mu_3\mu_4}}$. There is a natural extension of this rule to all terms in (\ref{WZact}) giving rise to the conjectured derivative corrections to the Born-Infeld part of the action:  
\be \label{prescr}
-T_9 \int \!\!\D^{10}x\exp [ \sum_{n\ge2} \ts{\frac{\ze(n)}{n}} \tr(\al'\mathsf{S}_{\mu\nu}\frac{\de}{\de F_{\nu\mu}} )^{n} ] \sqrt{h}\,.
\ee
If this prescription would continue to hold beyond $\cO(\al'^2)$, one could use it to obtain derivative corrections to the Born-Infeld action from the corrections to the Wess-Zumino term (above we have only given the prescription for the known derivative corrections to the Wess-Zumino term). This would be a powerful tool, since the corrections to the Wess-Zumino term are easier to calculate. In the rest of this section we will investigate whether such a relation could be true. 

First of all, the prescription as we have stated it (\ref{prescr}) is not precise. For instance, one has to decide how to order the $\frac{\de}{\de F}$'s vs.\ the $\mathsf{S}$'s when expanding the exponential. 
Another issue is that the expansion of the exponential terminates for the Wess-Zumino term since forms of degree higher than ten vanish identically, whereas, at least at first sight, this is not the case for the expression arising after using the above replacement rule. 
Setting these ambiguities aside for the moment, is there any way one can hope to test the proposed relation? As pointed out above it is in general very involved to calculate derivative corrections directly. 
Fortunately there is, however, at least one class of corrections that are within reach, namely the corrections which involve only powers of terms of the form $h_{A}^{\mu\nu}\pa_{\rho_1}\cdots \pa_{\rho_k}F_{\mu\nu}$ for the Born-Infeld term and wedge products of two-forms of the form $ \pa_{\rho_1}\cdots \pa_{\rho_k} F$ in the Wess-Zumino term. 

At this point we need to recall some terminology and formul\ae{} from \cite{Wyllard:2000}. We will be brief; the interested reader may refer to \cite{Wyllard:2000} and references therein for further details. The boundary state for a D9-brane in the presence of a general gauge field is \cite{Callan:1988}
\be \label{BFX}
|B(F(X))\rangle=e^{-\frac{i}{2\pi \al'}\int
d\si \left[ \pa_{\si}X^{\mu}A_{\mu}(X) -
\frac{1}{2}\Psi^{\mu}\Psi^{\nu}F_{\mu\nu}(X)\right] }\B.
\ee 
One can write this expression more compactly, by packaging $X^\mu$ and $\Psi^\mu$ into the superfield $\phi^{\mu} = X^{\mu} +
\tha\Psi^{\mu}$ and introducing the super-covariant derivative $D = 
\tha\pa_{\si} -\pa_{\tha}$, as $|B(F(X))\rangle=e^{-\frac{i}{2\pi \al'}\int
d\si \D\tha D \phi^{\mu}A_{\mu}(\phi) }\B$.  
It is known \cite{Callan:1988} that the parity-conserving contribution
to the effective action is proportional to  (in the absence of a $B_{\mu\nu}$-field) 
$S_{\mathrm{BI}} = \langle 0 |B(F(X))\rangle_{\mathrm{NS}}$, and that the
parity-violating part of the effective action which couples linearly
to the background RR form fields is obtained by
calculating the overlap between the state $|C\rangle$ representing these 
fields and the boundary state (see e.g.\ \cite{DiVecchia:1999c}),
i.e.\ $S_{\mathrm{WZ}} = \langle C|B(F(X)\rangle_{\mathrm{R}}$.

To study derivative corrections to the effective action one 
Taylor expands $A_{\mu}(\phi)$ around the zero mode of $\phi$ (which we denote $\vphi_0$), i.e.\ one splits $\phi$ as $\vphi_0 +
\tilde{\phi}$:  
\be \label{Aexp}
A_{\mu}(\phi) = \sum_{n=0}^{\infty}
\frac{1}{n!}\tilde{\phi}^{\nu_1}\cdots\tilde{\phi}^{\nu_n} \pa_{\nu_1} \cdots \pa_{\nu_n} A_{\mu}(\vphi_0) 
\ee
Using this result one finds
\bea \label{Fexp}
&&\!\!\!\!\!\!\!\!\!\int \D \si \D \tha D_{\si}\phi^{\mu}A_{\mu}(\phi) =\int \D \si \D\tha 
\sum_{k=0}^{\infty} \ts{\frac{1}{(k{+}1)!}\frac{k{+}1}{k{+}2}}D \tphi^{\nu}\tphi^{\mu}\tphi^{\la_1} \cdots
\tphi^{\la_k}\pa_{\la_1} \cdots \pa_{\la_k} F_{\mu\nu}(x) \non \\&& \qquad\qquad\qquad \quad - \int \D
\si [2\tilde{\Psi}^{\mu}\psi_0^{\nu} + \half\psi_0^{\mu}\psi_0^{\nu}] \sum_{k=0}^{\infty} \frac{1}{k!}\tX^{\la_1} \cdots
\tX^{\la_k}\pa_{\la_1} \cdots \pa_{\la_k} F_{\mu\nu}(x)
\eea
where we have used that in the NS sector $\vphi^{\mu}_0 = x^{\mu}$ whereas in the R sector
 $\vphi^{\mu}_0 = x^{\mu} + \tha \psi_0$ (thus, in the NS sector the second term in (\ref{Fexp}) is absent).

After this brief digression we now return to the discussion of the restricted class of derivative corrections described above. Let us start by calculating the relevant corrections to the WZ term. To this end, we keep only the terms of the form $\tX^k\psi_0^\mu \psi_0^{\nu} \pa^k F_{\mu\nu}$ in the expansion (\ref{Fexp}). This leads to the following contribution to the WZ term
\be
\langle C|e^{ \frac{i}{4\pi \al'} \psi_0^{\mu} \psi_0^{\nu} \int \D \si \sum_{k=2}^{\infty} \frac{1}{k!} \tX^{\rho_1} \cdots \tX^{\rho_k}\pa_{\rho_1}\cdots \pa_{\rho_k} F_{\mu\nu} }\BF_{\mathrm{R}}\,.
\ee 
Next we evaluate the zero-mode part using the result (here $|0\rangle$ denotes the vacuum state which is annihilated by all the annihilation oscillators)  
\be 
\langle C| (\psi_0^{\mu}\psi_0^{\nu}N_{\mu\nu})^{k} \BF_{\mathrm{R}} \rar  C\we \langle 0|(-2i\al'N)^k|B(F)\rangle_{\mathrm{R}}\,,
\ee
to obtain
 \be \label{WZres}
C \we \langle 0 |e^{ \frac{1}{2\pi} \int \D \si \sum_{k=2}^{\infty} \frac{1}{k!} \tX^{\rho_1} \cdots \tX^{\rho_k}\pa_{\rho_1}\cdots \pa_{\rho_k} F }\BF_{\mathrm{R}}\,.
\ee 
To proceed further one needs to calculate the various correlations functions of the form $\langle 0 |\tilde{X}^K \BF_\mathrm{R}$ which appear when expanding the exponential. These correlation functions can be expressed in terms of sums as in \cite{Wyllard:2000}, however, it appears to be very difficult to calculate all contributions to (\ref{WZres}) that way. We will circumvent this problem by applying the prescription directly to the expression (\ref{WZres}). This means that one needs to check separately that the renormalization works in the same way for the two expressions. 

By applying the above suggested rule to convert the expression (\ref{WZres}) into conjectured corrections to the Born-Infeld term, one finds
\be
-\langle 0 | e^{ \frac{1}{2\pi} \int \D \si \sum_{k=2}^{\infty} \frac{1}{k!} X^{\rho_1} \cdots X^{\rho_k}\pa_{\rho_1}\cdots \pa_{\rho_k} F_{\mu\nu}\frac{\de}{\de F_{\nu\mu}} }\BF_{\mathrm{R}} \sqrt{h}\,.
\ee 
Finally using the result $\frac{\de}{\de F_{\mu\nu}} \sqrt{h} = - \frac{1}{2} h_{A}^{\mu\nu} \sqrt{h}$ we find
\be \label{Prerel1}
-\langle 0 | e^{ -\frac{1}{4\pi} \int \D \si \sum_{k=2}^{\infty} \frac{1}{k!} X^{\rho_1} \cdots X^{\rho_k}h_{A}^{\mu\nu}\pa_{\rho_1}\cdots \pa_{\rho_k} F_{\mu\nu} }\BF_{\mathrm{R}} \sqrt{h} + \mbox{terms with other structures}\,.
\ee
This conjectured expression should be compared to the one obtained by direct calculation. 
To determine the relevant corrections to the Born-Infeld part we need to consider the expression $ \langle 0 |e^{-\frac{i}{2\pi\al'} \int \D \si \D \theta D \phi^{\mu} A_{\mu}(\phi) } \BF_{\mathrm{NS}}$ with the exponent expanded as in (\ref{Fexp}). We are only interested in the terms which have one internal contraction between $D\phi$ and a $\phi$. For such terms we find
\bea
&& D \phi^\nu \phi^\mu \phi^{\rho_1} \cdots \phi^{\rho_k} \pa_{\la_1} \cdots \pa_{\la_k} F_{\mu\nu} = \non \\ && [D_1 G^{\nu\mu}_{12}]_{2\rar 1} \phi^{\rho_1} \cdots \phi^{\rho_k} + k [D_1 G^{\nu\rho_1}_{12}]_{2\rar 1} \phi^{\mu}\phi^{\rho_2} \cdots \phi^{\rho_k} \pa_{\la_1} \cdots \pa_{\la_k} F_{\mu\nu}  = \non \\&&  {\ts \frac{1}{2} }\tha (i\al')(k+2)h_{A}^{\nu\mu} \phi^{\rho_1} \cdots \phi^{\la_k} \pa_{\la_1} \cdots \pa_{\la_k} F_{\mu\nu} \,,
\eea
where in the last step we have used the Bianchi identity and the result $[D_1 G^{\nu\mu}_{12}]_{2\rar 1} =  {\ts \frac{1}{2} }\tha (i\al')h_{A}^{\nu\mu}$. Using this result together with (\ref{Fexp}) we find
\be \label{Prerel2}
\langle 0 | e^{ -\frac{1}{4\pi} \int \D \si \sum_{k=2}^{\infty} \frac{1}{k!} X^{\rho_1} \cdots X^{\rho_k}h_{A}^{\mu\nu}\pa_{\rho_1}\cdots \pa_{\rho_k} F_{\mu\nu} }\BF_{\mathrm{NS}} \,,
\ee
which is almost the same as (\ref{Prerel1}), except that this expression does not have a factor of $\sqrt{h}$ and involves the NS sector whereas (\ref{Prerel1}) involves the R sector. However, since neither expression involves any explict fermions, the only difference will be in the overall $\langle 0 \BF$ factor. Using, $\langle 0 \BF_{\mathrm{R}} = T_p$, whereas $\langle 0 \BF_{\mathrm{NS}} = - T_p \sqrt{h}$, we find agreement.  One can argue that a renormalization of $F$: $F\rar F + \de F$ gives rise to the same change in the Wess-Zumino and Born-Infeld terms. For instance, the leading term arises from  $\de e^{F} = \de F e^F$ in the Wess-Zumino term whereas it arises from $\de F_{\mu\nu} \frac{\de}{\de F_{\mu\nu}} \sqrt{h}$ in the Born-Infeld term; these two changes are connected via the prescription.

We have thus shown that the prescription (\ref{prescr}) makes sense for an infinite class of terms. A few comments are in order. For consistency one has to formally extend the sum arising from the Wess-Zumino term to run over all values. For the terms
 we considered it was obvious how to do this; for more general corrections this might be ambgious. Furthermore, when expanding the exponential in (\ref{prescr}) all $\frac{\de}{\de F}$'s were brought to the right. Within the calculational scheme we have used the prescription appears to be natural. If another scheme is used to calculate derivative corrections integrations by parts might be needed to make the prescription work.

Is it possible to gather further evidence in favor of (\ref{prescr})? There is another class of corrections which are also amenable to analysis, namely terms which are built out of scalars involving four derivatives each. Writing the part of $|B(F(X))\rangle$ keeping only terms with at most two derivatives in (\ref{Fexp}) schematically as $\exp[\pa F + \pa^2F]|B(F)\rangle$ we find the following $4k$-derivative corrections to the Born-Infeld term keeping only the relevant contractions (we use a very schematic notation; the details are not important):   
\be 
{\ts \frac{1}{2k!} } \langle 0| (\pa^2 F)^{2k}|B(F)\rangle  \rar {\ts \frac{(2k-1)!!}{2k!} } \langle [\pa^2 F \pa^2 F]^k\rangle  = {\ts \frac{1}{k!} }\langle [\frac{1}{2} \pa^2 F \pa^2 F ]^k \rangle
\ee
 where $\langle \cdot \rangle$ is short hand for $\langle 0| \cdot |B\rangle_{\rm NS}$ and the terms inside square brackets are assumed to have all indices contracted. Similarly, one finds 
\be
{\ts \frac{1}{k!2k!} }\langle(\pa^2 F)^k(\pa F)^{2k}\rangle \rar {\ts \frac{1}{k!2k!} \frac{2k!}{2^k} }\langle [\pa^2 F \pa F \pa F]^k \rangle  = {\ts \frac{1}{k!} }\langle [\frac{1}{2} \pa^2 F \pa F \pa F]^k \rangle 
\ee 
and also $\frac{1}{4k!} (\pa F)^{4k}) \rar \frac{1}{4k!}\frac{4k!}{(4!)^k k!} \langle[ (\pa F)^4 ]^k\rangle = \frac{1}{k!} \langle [\frac{1}{4!} (\pa F)^4 ]^k \rangle $. From these results it follows that the corrections to the Born-Infeld term can be written $-T_p\sqrt{h}\frac{1}{k!}(L_4)^k$, where $L_4$ is equal to the four-derivative part of the action (\ref{BIact}). 
Let us now turn to the Wess-Zumino term. It is clear that only the $\ze(2)$ term contributes since all the other terms involve contractions between more than two $\pa^n F$'s (this is also necessarily true for all the corrections that we have not determined as one can convince oneself by using the results in \cite{Wyllard:2000}). Applying the prescription we get $-T_p\frac{1}{k!}(\frac{\ze(2)}{2}\al'^2 \tr(S)^2)^k (\frac{\de}{\de F})^{2k} \sqrt{h} = -T_p\frac{1}{k!}(L_4)^k\sqrt{h} + \mbox{other terms}$. Here we have used the fact that there is only one way in which the $\frac{\de}{\de F}$'s can act to give terms of the required form. We thus find agreement also for this class of terms.

Above we have gathered some evidence for the correctness of the prescription (\ref{prescr}). There are of course other corrections to the Wess-Zumino term, which will modify or invalidate this prescription. For instance, there are also six-derivative four-form corrections to the Wess-Zumino term (the two-form six-derivative corrections can be removed by field redefinitions). 
Also, there might be additional corrections arising from the expansion of the form fields $C$ as $C(X) = C(x) + X^{\mu}\pa_{\mu}C(x) + \ldots$.  Roughly speaking, it might be that all these additional contributions involve covariant derivatives of $S$ and there might be a sense in which they can be neglected.  A very non-trivial check of the ``all order in $S$'' proposal (\ref{prescr}) would be to check that it is invariant under the SW map. 
Unfortunately, the method used in this paper makes this rather involved since several integrations by parts are needed to make the invariance manifest. The approach pursued in \cite{Das:2001} looks more promising, but there is of course no guarantee that the restriction to  only the corrections (\ref{WZact}) leads to corrections to the Born-Infeld action that are invariant by themselves. Finally, as an example we will write down explicity the form of the six-derivative corrections that arise from the prescription: 
\bea
&&-T_p \int \D^{10}x \sqrt{h} {\ts \al'^3\frac{\ze(3)}{6}} \Big[ \Big\{ h^{\mu_2 \mu_3 }h^{\mu_4 \mu_5}h^{\mu_6 \mu_1} + h^{\mu_2 \mu_5 }h^{\mu_6 \mu_3}h^{\mu_4 \mu_1} + \half h^{\mu_5 \mu_6 }h^{\mu_4 \mu_1}h^{\mu_2 \mu_3} \non \\ 
&& \qquad \qquad \qquad \qquad +\,\half h^{\mu_3 \mu_4 }h^{\mu_2 \mu_5}h^{\mu_6 \mu_1} +\half h^{\mu_1 \mu_2 }h^{\mu_4 \mu_5}h^{\mu_6 \mu_3}  -(\half)^2 h^{\mu_1 \mu_2 }h^{\mu_3 \mu_4}h^{\mu_5 \mu_6} \Big\} \non \\  
&& \qquad \qquad \qquad \qquad \quad \times\, h^{\rho_2 \rho_3} h^{\rho_4\rho_2} h^{\rho_6\rho_1} S_{\rho_1\rho_2\mu_1\mu_2} S_{\rho_3\rho_4\mu_3\mu_4} S_{\rho_5\rho_6\mu_5\mu_6} \Big]\,.
\eea
A first check of these terms would be to check that they vanish in the Seiberg-Witten limit used in \cite{Das:2001}, as this is required for consistency with the approach used there. 

\begingroup\raggedright\endgroup

\end{document}